\documentclass[doublecol]{epl2}
\usepackage{amsmath}
\usepackage{amssymb}
\usepackage{mymacros}
\usepackage{epsfig}% Include figure files
\usepackage{graphicx}
\usepackage{epsfig}
\usepackage{amsfonts}

\title{The Valence Bond Glass phase}


\author{Marco Tarzia and Giulio Biroli}
\institute{Institut de Physique Th{\'e}orique, Orme des Merisiers -- CEA Saclay, 91191 Gif sur Yvette Cedex, France}

\abstract{We show that a new glassy phase can emerge
in presence of strong magnetic frustration and quantum fluctuations.
It is a Valence Bond Glass. We study its properties solving the 
Hubbard-Heisenberg model on a Bethe lattice within the large $N$ limit 
introduced by Affleck and Marston.
We work out the phase diagram that contains Fermi liquid, dimer and
valence bond glass phases. This new glassy phase
has no electronic or spin gap (although a pseudo-gap is observed),
it is characterized by long-range critical valence bond correlations
and is not related to any magnetic ordering. As a consequence it is quite
different from both valence bond crystals and spin glasses.}

\begin{document}

\maketitle
The interplay of strong quantum fluctuations and geometrically 
frustrated magnetic interactions can give rise to new 
low temperature phases. As noticed by Anderson~\cite{Anderson1} 
a way to minimize the effect of frustration and obtain a low energy
state is coupling the electrons in valence bonds. A very good 
variational wave function that is generically in competition with the 
antiferromagnetic (or more general magnetic) state can be obtained by forming 
a superposition of short range valence bonds 
that are arranged as dimers on the lattice.  
If no lattice symmetry is broken this corresponds to the (so
 called) resonating valence 
bond liquid (RVBL). In the last decades, this state has received a lot of 
attention in connection with the unusual physical
behavior of the normal phase of underdoped high $T_c$ 
superconductors~\cite{Lee}. Indeed Anderson~\cite{Anderson2} proposed that the holes 
created by doping the antiferromagnetic insulator (of the high $T_c$'s phase diagram) can gain 
substantial kinetic energy in the RVBL state and not in an antiferromagnetic
background. As a consequence, doping favors the RVBL state which 
could then become the thermodynamic stable phase and be 
responsible for the unusual behavior of 
underdoped samples. Concomitantly, resonating valence bond ground states 
have been the focus of an intense activity~\cite{Misguich} in the context
of frustrated magnets. RVBL or spin 
liquids have been found for several models~\cite{Misguich}. 
These states can undergo quantum phase transitions where  
lattice symmetries are spontaneously broken. This gives rise 
to valence bond {\it crystals} (VBC). 
Different models are known to lead to this 
type of ground states~\cite{Misguich} characterized by long range dimer-dimer 
correlations. The situation in experiments is 
complicated by unavoidable magneto-elastic couplings: making the 
difference between induced and spontaneous dimerization is a difficult task. 
A first experimental example of spontaneously broken states 
has been apparently found in~\cite{Tamura}.               \\
The aim of this work is to study a new kind of valence bond state: 
the valence bond glass (VBG). Similarly to VBC the arrangement of 
the dimers (or valence bonds) 
breaks the lattice symmetry. However, contrary to VBC, 
this corresponds to an amorphous dimerization and not crystalline one.
%in crystalline arrangements. The resulting 
%configuration is characterized by long range dimer-dimer correlations: 
%it is a VB solid that shows a chaotic (or disordered) dimerization. 
Although VBG are analogous to spin glasses~\cite{SG}
they are physically quite different. In particular 
the spins do not freeze in a disordered profile. 
We expect that the VBG phase can arise in presence of strong 
magnetic frustration as one of the competing ground states. 
The addition of (little) quenched disorder will favor this phase.
Depending on the system, the low temperature phase could be either a
VBG or a spin glass. Actually, the spin glass phase is conjectured to exist 
even in absence of disorder on some frustrated lattices~\cite{Expt1} 
(see however~\cite{Henley,Expt2}). \\
In the following we shall investigate the properties of the valence bond glass phase 
focusing on the Hubbard-Heisenberg model within the large $N$ approximation 
introduced by Affleck and Marston~\cite{affleck}. The underlying lattice 
we shall focus on is a random regular graph with connectivity $z$\footnote{
It is a graph taken at random within the set of graphs whose $\cal N$ sites are all connected to $z$ randomly choosen neighbors.}. The reason for this choice is twofold. First,
this type of graphs are on any finite lengthscale as Bethe lattices or Cayley trees. This, as it is well known 
for classical systems~\cite{baxter}, introduces useful simplification in the
analysis of the model. The main reason is, however, that topological 
frustration and quenched disorder are introduced by very long loops 
(of the order $\log {\cal N}$  where ${\cal N}$ is the number of sites) in 
random regular graphs.  These loops disfavor crystalline 
states and let emerge easily the 
glassy phases~\cite{ParisiMezard,BiroliMezard}.    
We consider the $SU(N)$ version of the  
Hubbard-Heisenberg model introduced in~\cite{affleck}:
\begin{eqnarray} \label{eq:ham}
\nonumber
{\cal H} &=& -t \sum_{\langle i, j \rangle} \left( c^{\dagger}_{i, \alpha} 
c_{j,\alpha} + \textrm{h.c.} \right) + \frac{U}{N} \sum_i \left ( n_i - 
\frac{N}{2}
\right)^2 \\ 
&& \qquad + \, \frac{J}{N}  \sum_{\langle i, j \rangle} \mathbf{S}_i \cdot 
\mathbf{S}_j,
\end{eqnarray}
where $c_{i, \alpha}$ denotes the destruction operator of an electron of
spin index $\alpha$ ($\alpha = 1, \ldots, N$ with $N$ even) on the site 
$i$. The sum $\langle i, j \rangle$ is restricted on nearest neighbor sites 
on the lattice. The first two terms correspond to the $SU(N)$ Hubbard model, 
while the last term accounts for the nearest neighbor 
antiferromagnetic interaction ($J>0$)\footnote{As discussed in~\cite{affleck}, the 
antiferromagnetic interaction is not generated in perturbation theory at $N=\infty$,
so it has to be added in the original Hamiltonian.}. 
We shall focus on the $N\rightarrow \infty$ limit and 
consider only the half-filling case, where
$n_i /(N/2)= \sum_{\alpha}  c^{\dagger}_{i, \alpha} 
c_{i,\alpha}/(N/2)=1$ for all sites. 
%in other words, we consider the $SU(N)$ $t$-$J$
%model at half filling). 
Using that 
${\bf S}_i \cdot {\bf S}_j$ equals $-\sum_{\alpha,\beta } c^{\dagger}_{i, \alpha} 
c_{j,\alpha} c^{\dagger}_{j, \beta} c_{i,\beta}$ 
up to constant terms in the large $N$ limit~\cite{affleck,rokhsar}, 
the Hamiltonian can be rewritten in a $SU(N)$ manifestly invariant
form. At half-filling it reads:
\begin{equation}
{\cal H} = -t \sum_{\langle i, j \rangle} \left( c^{\dagger}_{i, \alpha} c_{j,\alpha} 
+ \textrm{h.c.} 
\right ) - \frac{J}{N} \sum_{\langle i, j \rangle} c^{\dagger}_{i, \alpha} 
c_{j,\alpha} c^{\dagger}_{j, \beta} c_{i,\beta}.
\end{equation}
Note that all terms constant in the large $N$ limit have been neglected.
Here and henceforth the summation over the $SU(N)$ indices will be skipped for simplicity.
%Within the finite temperature formalism, 
The partition function of the system at finite temperature
can be written as a path integral
\begin{equation} \label{eq:z}
Z = \int \! {\cal D} c \, {\cal D} c^{\dagger} \exp \left[ - \int_0^{\beta} \! 
\textrm{d} \tau \, {\cal L} (c, c^{\dagger}) \right],
\end{equation}
where $\beta$ is the inverse temperature, and the (imaginary time) Lagrangian
is ${\cal L} (c, c^{\dagger}) = {\cal H} + \sum_i c^{\dagger}_{i,\alpha} 
\left( \textrm{d} / \textrm{d} \tau \right) c_{i, \alpha}$. 
%Antiperiodic boundary conditions at time $\tau = 0$ and $\tau = \beta$ are 
%imposed on the Grassmanian fermionic operators.
The functional integral is of course non trivial, due to the presence of the
non linear interaction. However, one can perform a Hubbard-Stratonovich 
transformation which allows to rewrite the Lagrangian quadratically in the 
fermions, at the expense of introducing a new (complex) bosonic field, 
$\chi_{ij}$, on each edge of the lattice~\cite{affleck}:
%The new Lagrangian reads:
\begin{eqnarray}
\nonumber
{\cal L} (c, c^{\dagger}, \chi) &=& 
\sum_{\langle i, j \rangle} \left \{ \frac{N}{J} | \chi_{ij} |^2 -
\left[ \left( t + \chi_{ij} \right) c^{\dagger}_{i, \alpha} c_{j,\alpha} + 
\textrm{h.c.} \right ] \right \}\\
&& \qquad + \, \sum_i c^{\dagger}_{i,\alpha} \left( 
\frac{\textrm{d}}{\textrm{d} \tau} \right) c_{i, \alpha}.
\end{eqnarray}
The equation of motion of the auxiliary bosonic field reads:
\begin{equation} \label{eq:eom}
\langle \chi_{ij} (\tau) \rangle = \, 
\frac{J}{N} \langle c^{\dagger}_{j, \alpha} (\tau) c_{i,\alpha} (\tau) \rangle.
\end{equation}
$\chi_{ij}$ is the valence bond field and gives an extra contribution to the electron
hopping amplitude between the sites $i$ and $j$. The number of {\it valence bonds} 
on link $(ij)$ is given by $N| \chi_{ij} |^2/J$ up to subleading terms~\cite{affleck}.\\
%Valence bond states minimize the Heisenberg interaction:
%if one considers the action of the Hamiltonian on a valence bond,
%one notes that a $SU(N)$ singlet on a given link is an eigenstate of ${\cal H}$
%with the lowest possible eigenvalue ($-J$).
The advantage of this representation is that the integral over the fermionic 
degrees of freedom is now Gaussian. Therefore, they can 
be integrated out, leading to an effective action which depends only on
the bosonic variables:
\begin{equation}
\exp \left[ - S_{eff} (\chi) \right] = \int \! {\cal D} c \, 
{\cal D} c^{\dagger} \exp \left[ - \int_0^{\beta} \!\! \textrm{d} \tau \, 
{\cal L} (c, c^{\dagger}, \chi) \right].
\end{equation}
The effective action thus reads:
\begin{equation}
S_{eff} = N \int_0^{\beta} \!\! \textrm{d} \tau \sum_{\langle i, j 
\rangle} \frac{1}{J} | \chi_{ij} |^2 - N \textrm{Tr} \, \log {\mathbb M},
\end{equation}
where the matrix ${\mathbb M}$ is given by ${\mathbb M} = [ ( \textrm{d} / 
\textrm{d} \tau ) {\mathbb I} -t \mathbb{C}-\hat{\chi} ]$, 
$\mathbb{C}$ being the connectivity matrix of the lattice, i.e., $\mathbb{C}_{ij}=1$ if $i$ and $j$ are nearest neighbors on the lattice and zero otherwise.
$\hat{\chi}$ has an analogous definition except that $\hat{\chi}_{ij}=\chi_{ij}$ if $i$ and $j$  are nearest neighbors.\\
So far, these transformations are exact and do not depend on the
particular choice of the lattice. In the $N \to \infty$ limit the
saddle point integration over the bosonic variables, $\chi_{ij}$, 
becomes exact and we can compute the free energy of the system by  
seeking the lowest minimum of the effective action\footnote{If we had decoupled the $U$ term 
 in eq. \ref{eq:ham}, as done for the $J$ term, by introducing a field $\phi_i$ then we would have found saddle point equations leading, at half filling, to the solution $\phi_i=0$~\cite{affleck}. That is
the reason why we dropped this term from the beginning.}. 
Assuming that at the saddle point the valence bond operators are time-independent, the problem
reduces to finding the minima of the ``classical'' free energy $\beta F(\chi) = 
S_{eff}/N$ (N being the number of SU(N) indices),
\begin{equation} \label{eq:free}
F(\chi) = \sum_{\langle i, j \rangle} \frac{1}{J} | \chi_{ij} |^2 - 
\frac{1}{\beta} 
%\int \textrm{d} \lambda \, \rho (\lambda) \,
\sum_{\lambda} \log \left [ 1 + \exp \left( -\beta \lambda \right) \right] \quad.
\end{equation}
We denote by $\lambda$ the eigenvalues of the one-particle Hamiltonian
\begin{equation} \label{eq:h1}
{\cal H}_1 = -\sum_{\langle i, j \rangle} \left[ \left( t + \chi_{ij} \right) 
c^{\dagger}_i c_j + \textrm{h.c.} \right ] \quad.
\end{equation}
Note that the (complex) bosonic variables $\chi_{ij}$ can have any 
arbitrary spatial dependence and 
that there is no need to introduce the chemical potential since
it is expected, and found, to be zero at half filling\footnote{Although random regular graphs
are not bipartite, they behave in a similar way. In particular, for all phases, 
we find electronic 
densities of state that are symmetric around zero. Thus, the chemical potential is zero at half filling.}. 
For simplicity we will set $J=1$ in the following, 
bearing in mind that all energy scales are measured in units of $J$.\\
The saddle point 
equations consist simply in Eq.~(\ref{eq:eom}) where the average on the RHS is
performed using the Hamiltonian ${\cal H}_1 $. Obtaining an analytical solution 
for a given particular lattice is, in general, a hard task. 
However, in some special cases, the problem
can be simplified.
%the nature of the 
%configurations of the valence bonds which minimize the effective action.
In particular by considering periodic solutions one reduces 
the independent degrees of freedom 
to a finite number ($4$ in the case studied by Affleck and Marston~\cite{affleck}).
Our aim is to find whether there are amorphous or chaotic solutions. 
Thus, in our case, obtaining a full analytical solution seems extremely difficult.\\
On infinite random graphs the Bethe-Peierls 
approximation is exact \cite{ParisiMezard}:
since the average length of the loops is infinite,   
it is possible to write down
self-consistent iteration equations for local ``cavity'' Green's functions,
(or ``Weiss functions''), ${\cal G}_i$, defined on each site of the 
graph~\cite{rmp}.    
In particular, for any given configuration of the valence bonds, $\{ 
\chi_{ij} \}$,
it is straightforward to show that the following recursion relations must
hold:
\begin{equation} \label{eq:cav1}
{\cal G}_i (\nu_n) = i \nu_n - \sum_j^{z-1}
\frac{ | t + \chi_{ij} |^2 }{{\cal G}_j (\nu_n)},
\end{equation}
where $\nu_n = (2n+1) \pi / \beta$ are the fermionic Matsubara 
frequencies. The Green's function, $G_i (\nu_n) = - \beta 
\langle c_{i \alpha} (\nu_n) 
c_{i \alpha}^{\dagger} (\nu_n) \rangle$, can be calculated on each site  
as a function of the ${\cal G}_i$ on the neighboring sites, by using 
Eq.~(\ref{eq:cav1}), where the sum is extended over all the $z$ neighbors.
For any given finite graph, and for any given profile of the bosonic field,
Eqs.~(\ref{eq:cav1})
provide a set of solvable equations for the cavity propagators.
%In the context of spin glasses these equations are called {\em belief
%propagation} equations, and could, in principle, be solved for any
%individual instance of the graph and $\{ \chi_{ij}\}$.
Furthermore, by enforcing the equation of motion for the valence bonds, 
Eq.~(\ref{eq:eom}), one finds that, on each link of the graph, 
the bosonic operators must verify:
\begin{equation} \label{eq:cav2}
\chi_{ij} = 
%- \, \frac{J}{\beta^2} \sum_n \langle c_i^{\dagger} (\nu_n) c_j (\nu_n) 
%\rangle = 
 - \, \frac{1}{\beta} \sum_n \frac{t + \chi_{ij}}
{{\cal G}_i (\nu_n) {\cal G}_j (\nu_n) - | t + \chi_{ij}|^2}.
\end{equation} 
The last equation is non-local, and is 
reminiscent of the TAP equations derived in the context
of spin glasses~\cite{TAP}.
For infinite systems 
Eqs.~(\ref{eq:cav1}) and (\ref{eq:cav2}) allow to treat the
liquid and the dimer phase (see below) in a very natural way. 
The analysis in the glass phase is much more involved and complicated. 
See~\cite{ParisiMezard} for the method used in classical cases\footnote{
The cavity method that would be needed to analyze the glassy phase 
is substantially more difficult than the one developed for spin glasses
on Bethe lattices. The reason is that 
the valence bond interaction is on all scales and not only 
between nearest neighbors.} and~\cite{Sondhi} for its extension to 
quantum cases.
As a consequence we will use the previous approach to study simple (non disordered) phases 
and the transition lines. In order to study the glassy phase we
interpret the free energy, 
Eq.~(\ref{eq:free}), as the Hamiltonian of a classical system of complex 
variables. Hence, the problem of finding the minima of the free energy
is reduced to finding classical ground states. 
%of a {\it classical} system with Hamiltonian  given by eq.~(\ref{eq:free}). 
To solve the latter problem we use Monte Carlo 
annealing simulations. Basically, we introduce an auxiliary temperature $T_{aux}$ and, at each step, 
we attempt to change one $\chi_{ij}$ at random according to the 
Boltzmann weight $e^{- F(\chi)/T_{aux}}$.
The move is accepted with
probability $ p = \textrm{min} \left \{ 1, \exp [ -\Delta F / 
T_{aux}] \right \}$. The auxiliary temperature is finally decreased
at constant rate down to zero temperature. Details on the numerical procedure
are discussed in the Appendix.\\ 
\begin{figure}
\begin{center}
\includegraphics[scale=0.29,angle=270]{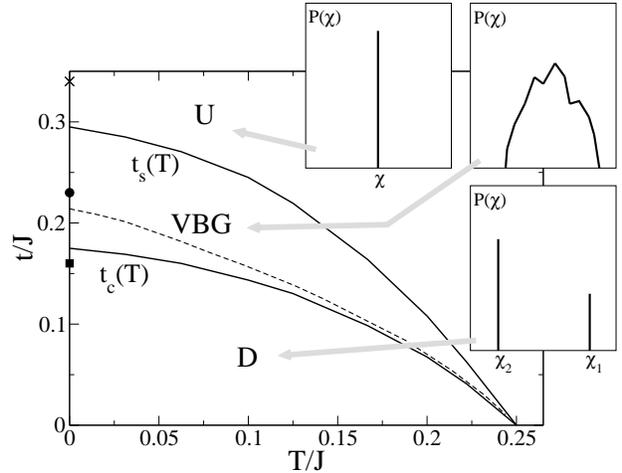}
\end{center}
\vspace{-0.35cm}
\caption{Phase diagram of the Hubbard-Heisenberg $SU(N)$ model 
at half filling on the
random regular graph ($z=3$). We show the relative positions of the
uniform phase (U), the dimer phase (D), and the valence bond glass (VBG).
At $t_s(T)$ the uniform phase becomes unstable, the valence bond
non-linear susceptibility diverges (see Fig.~\ref{fig:susc}), and a
continuous transition from the Fermi liquid to the VBG takes place. At 
$t_c(T)$ the free energies of the dimer phase and that of the VBG coincide and
a first-order transition occurs. The dashed line corresponds to the spinodal
of the dimer phase. The probability distributions of the
valence bonds in the different phases are reproduced schematically in the
insets.} 
\label{fig:pd}
\vspace{-0.35cm}
\end{figure}
By employing both the analytical and the numerical approaches described
above, we have derived the phase diagram of the $SU(N)$
Hubbard-Heisenberg model on the random regular graph with connectivity $z=3$, 
see Fig.~\ref{fig:pd}.\\
%Such phase diagram is reported in Fig.~\ref{fig:pd}
%as a function of the temperature $T$ and the hopping parameter $t$,
%showing the relative position of the different phases, and the
%transition lines between them. \\
{\it Uniform phase---}At 
high enough temperature and hopping amplitude
the system is in a {\it uniform phase},
where the bond operators are real and equal on each link of the graph, 
$\chi_{ij} = \chi$. 
For a given value of $\chi$, the electronic density of states 
can be computed easily since the density of states 
of the connectivity matrix is known ~\cite{bray}, see the inset of Fig. 2. 
The uniform phase is translational invariant and gapless. It is 
clearly a Fermi liquid. \\
%In this case, the free energy (per site) 
%of the system reads:
%\begin{equation} \label{eq:funif}
%F = \frac{z \chi^2}{2J} - \frac{1}{\beta} \int \! \textrm{d} \lambda \,
%\rho(\lambda) \log \left [ 1 + e^{ - \beta \left( t + \chi \right)
%\lambda } \right ],
%\end{equation}
%where the electron spectrum, $\rho(\lambda)$, corresponds to the 
%spectrum of the connectivity matrix of the random graph, ${\mathbb C}$, 
%which can be
%computed exactly~\cite{bray}. 
%corresponds to a Fermi liquid.
For each value of $T$ and 
$t$, $\chi(T,t)$ in the uniform phase 
%is given by the minimization of the free energy, Eq.~(\ref{eq:funif}), 
%with respect to $\chi$.
%Equivalently, $\chi$ 
can be computed within the Bethe approximation,
by enforcing translational invariance into Eqs.~(\ref{eq:cav1}) and
(\ref{eq:cav2}) (i.e., ${\cal G}_i = {\cal G}$ and
$\chi_{ij} = \chi$), 
which reduce to a simple algebraic equation:
\begin{equation}
\chi = \sum_n \frac{(t + \chi)/\beta}
{\frac{\nu_n^2}{2} + (z-2)|t+\chi|^2 + \nu_n 
\sqrt{\frac{\nu_n^2}{4} + (z-1)|t+\chi|^2}}.
\end{equation}
One can then check the stability of the liquid solution with respect to
any other solution of the bosonic field. This
amounts in studying the (lowest) eigenvalues of the Hessian of $F(\chi)$. 
Using the base where the one-particle Hamiltonian, Eq.~(\ref{eq:h1}), is 
diagonal, and Fourier transforming with respect to the imaginary time,
one gets
%$(\mathbb{H})_{(ij),(kl)} (\tau - \tau^{\prime}) 
%= \partial^2 F / \partial \chi_{ij} (\tau)
%\partial \chi_{kl} (tau^{\prime})$. 
\begin{eqnarray} \label{eq:stab}
&&\!\!\!\!\!\!\!\!\!\!\!\!\!\!\!\frac{\partial^2 F (\chi)}
{\partial \chi_{ij}(\omega_n) \partial \chi^{\star}_{kl}(\omega_n)}  =
\frac{1}{J} \, \delta_{(i,j)(l,m)} 
- \sum_{\lambda, \lambda^{\prime}} v_{\lambda}^{i} v_{\lambda}^j
v_{\lambda^{\prime}}^{l} v_{\lambda^{\prime}}^k\\
\nonumber
&& \,\,\,\,\,\,\,\,\,\,\, 
\times \, \frac{1 - e^{\beta (\lambda + \lambda^{\prime})}}
{1 + e^{\beta \lambda} + e^{\beta \lambda^{\prime}} +
e^{\beta (\lambda + \lambda^{\prime})} } \,
\frac{\lambda + \lambda^{\prime}}{\omega_n^2 + 
\left( \lambda + \lambda^{\prime} \right)^2},
\end{eqnarray}
where $v_{\lambda}^{i}$ is the $i$-th component of the eigenvector
associated with the eigenvalue $\lambda$, and $\omega_n = 2 n \pi / \beta $ are
the bosonic Matsubara frequencies. 
The first instability of the uniform solution is expected to correspond
to a long wave-length modulation and should thus occur at $\omega_{n=0}$
first. In order to analyse it, we generate random regular graphs of size 
${\cal N}$ 
and compute $\lambda, v^i_{\lambda}$. Then using Eq.~(\ref{eq:stab}), we find that  
the smallest eigenvalue of the Hessian matrix at zero frequency 
becomes negative as either
$T$ or $t$ are decreased down to $t_s(T, {\cal N})$.
We then extrapolate the value of $t_s (T, {\cal N})$ (averaged over several
realisations of the graph) in the 
${\cal N} \to \infty$ limit by increasing ${\cal N}$ from $64$ to $1024$.
The curve $t_s(T)$ in the thermodynamic limit is shown in Fig.~\ref{fig:pd}. 
In particular, at $T=0$ the liquid solution becomes unstable at
$t_s \simeq 0.29$.\\
%As a consequence, the uniform solution becomes unstable
%with respect to other solutions with a spatial dependence of the bosonic field.
{\it Dimer phase---}At low enough temperature and hopping amplitude
a {\it dimer phase} (or {\em Peierls phase})~\cite{affleck}
is found to minimize the system free energy. In this
phase the valence bonds can assume only two possible 
values, $\chi_1$ on
${\cal N}/2$ links and $\chi_2$ on the others ${\cal N}(z-1)/2$, 
with $|\chi_1|>|\chi_2|$, 
in such a way that each site has exactly one link where the bosonic
operator equals $\chi_1$ and $z-1$ links where it equals $\chi_2$. 
As the random regular graph is dimerizable~\cite{matching}, the analysis  
of Ref.~\cite{rokhsar} guarantees that a dimer phase 
(with $\chi_2=0$) is the actual ground state of the pure
antiferromagnetic system ($t=0$).\\
At any given temperature and hopping amplitude, 
$\chi_1$ and $\chi_2$ can be determined analytically within the Bethe 
approximation. 
More precisely, one allows the cavity Green's functions and the valence
bonds to assume only two possible values, respectively ${\cal G}_1$ and
${\cal G}_2$, and $\chi_1$ and $\chi_2$. Taking into account the structure 
of the dimerized configurations, one can obtain a closed set of equations,
which can be easily solved:
\begin{eqnarray}
\nonumber
{\cal G}_a (\nu_n) &=& i \nu_n -
\left \{
\begin{array}{ll}
(z-2) \frac{ | t + \chi_2 |^2 }{{\cal G}_1 (\nu_n)} + 
\frac{ | t + \chi_1 |^2 }{{\cal G}_2 (\nu_n)} & \textrm{if~} a=1\\
(z-1) \frac{ | t + \chi_2 |^2 }{{\cal G}_1 (\nu_n)} 
& \textrm{if~} a=2
\end{array}
\right.\\
\chi_{a(b)} & = & 
- \, \frac{1}{\beta} \sum_n \frac{t + \chi_{a(b)}}
{\left[ {\cal G}_{b(a)} (\nu_n) \right]^2 - | t + \chi_{a(b)}|^2}.
\end{eqnarray}
In the dimer phase, both $\chi_1$ and $\chi_2$ turn out to be  
real (but at $t=0$, where the
system has a local gauge symmetry, $c_{i \alpha} 
\to c_{i \alpha} e^{i \theta_i}$ and
$c_{i \alpha}^{\dagger} \to c_{i \alpha}^{\dagger} 
e^{-i \theta_i}$). The electron spectrum in the dimer 
phase can be found similarly by computing the resolvent of
the matrix $t \mathbb{C}+\hat{\chi}$ in the dimerized state. 
The (electronic) density of state has gap, see
inset of Fig.2. This also induces a gap in the spin excitations\footnote{The spin Green function can be obtained quite easily from the electron Green function
in the large $N$ limit~\cite{affleck}.}.  
Using the above results,
the free energy of the dimer phase can be determined exactly for each
value of $T$ and $t$. At small enough temperature and hopping 
amplitude the dimer phase corresponds to the absolute minimum of the 
free energy.
% the free energy of the valence bond crystal is lower than
%that of the uniform pahse. 
%As $t$ (or $T$) are increased, the free energy of the dimer phase crosses that of 
%the uniform phase
%(e.g., at $t/J\simeq0.205$ at zero temperature) where a first order transition
%from the uniform to the dimer phase take place.
For larger values of $t$ (or $T$) the dimer phase reaches the spinodal line, 
where the gap closes and the smallest eigenvalue of the free 
energy Hessian matrix vanishes (dashed line in Fig.~\ref{fig:pd}).  
At zero temperature this happens at $t\simeq 0.218$.
Note that this zero temperature spinodal point lies below the corresponding one of the 
liquid which is the stable phase at high $t$. As a consequence, there is necessarily 
an intermediate phase. As we shall show 
in the following this is the Valence Bond Glass.   \\
\begin{figure}
\begin{center}
\includegraphics[scale=0.29,angle=270]{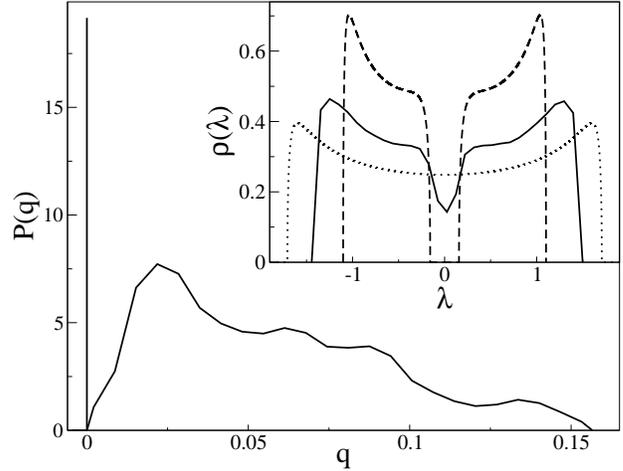}
\end{center}
\vspace{-0.35cm}
\caption{{\bf Main frame:} 
Overlap probability distribution, $P(q)$,
% = \overline{\sum_{a,b} w_{a}w_{b} 
%\delta (q - q_{ab})}$ 
at zero temperature and $t=0.23$ in the VBG. 
The data are averaged 
over $16$ different realizations of the graph, with ${\cal N} = 256$. 
The delta function in $q = 0$ corresponds to the fraction of 
replicas which end up in the same state, and it is expected to 
disappear in the thermodynamic limit (e.g., for a 
system of ${\cal N} = 128$ sites the delta peak in zero is approximately 
$1.5$ bigger than that for ${\cal N} = 256$). 
{\bf Inset:} Electron spectrum, $\rho (\lambda)$, 
at zero temperature in the different phases: 
Fermi liquid at $t=0.34$ (corresponding to the point marked by 
$\times$ in the phase 
diagram of Fig.~\ref{fig:pd}, dotted line), Valence Bond Glass at $t=0.23$ 
(point marked by $\bullet$ in Fig.~\ref{fig:pd}, 
continuous line) and Dimer phase at $t=0.16$ (point marked by 
$\blacksquare$ in 
Fig.~\ref{fig:pd}, dashed line). The electron 
spectrum has been computed analytically in the uniform and in the dimer phase, 
and numerically in the VBG phase.}
\label{fig:pq}
\vspace{-0.35cm}
\end{figure}
{\it Valence Bond Glass---} In order to study and prove the existence of the 
Valence Bond Glass phase we use Monte Carlo annealing simulations for the 
reasons explained previously. First, we check that our numerical procedure
gives back the uniform (dimer) phase at high (low) enough temperature
and hopping amplitude.
In the intermediate region where both phases are unstable
(e.g., at zero temperature for $0.218 < t \le 0.29$)
we find that amorphous configurations of $\chi_{ij}$
correspond to the actual minima of the free energy.
This is a glassy phase, which we call {\it valence bond glass}.
This is not a {\it spin} glass since
the average value of the spin is zero on
each site of the lattice, $\langle \mathbf{S}_i \rangle = 0$, as the
$SU(N)$ symmetry is unbroken.\\
%, and the spin-spin correlations, 
%$\langle {\mathbf S}_i \cdot {\mathbf S}_k \rangle_c^2$, 
%are short ranged. Conversely, long range spatial correlations
%among valence bonds on different links of the lattice, 
%$\langle \chi_{ij} \, \chi_{kl} \rangle_c^2$, emerge.
%Unfortunately, due to the fact that Eq.~(\ref{eq:cav2}) is 
%non-local, as it involves all the valence bonds of the graph,
%we are not able to describe the glassy phase analytically within
%the Bethe approximation. In order to do that, one should devise
%a more complicated formalism which takes into account the 
%existence of many glassy states, and the emergence of replica
%symmetry breaking~\cite{cavity}. 
%Therefore, our results on the VBG phase are mostly based on the
%MC numerical simulations.
The valence bonds, $\chi_{ij}$, are real valued 
and their disordered
profile is described by a nontrivial
distribution, $P(\chi)$, as schematically 
depicted in the inset of Fig.~\ref{fig:pd}. 
The electron spectrum is gapless in the VBG, 
although it exhibits a pseudo gap, as
shown in the inset of Fig.~\ref{fig:pq}, which becomes
deeper and deeper as either the temperature or the hopping amplitude
are decreased. \\
Interestingly enough, similarly to spin glasses~\cite{SG}, 
on any given graph 
different annealing procedures may lead to different configurations 
with the same free energy. 
One can measure the distributions of the overlaps between different states,
defined as: $
q_{ab} = \frac{2}{z{\cal N}} 
\sum_{\langle i, j \rangle} | \chi_{ij}^a - \chi_{ij}^b |$.
According to this definition, 
$q_{ab} = 0$ if the bosonic field has the same configuration in the two states,
whereas $q_{ab}>0$ otherwise. As in spin glasses, one can define 
the overlap distribution $P(q)=\sum_{a,b}w_aw_b \delta(q-q_{a,b})$ where 
$w_a$ is the thermodynamic weight of the amorphous state $a$~\cite{SG}. 
The overlap distribution is apparently continuous. 
$P(q)$, averaged over $16$ different realizations
of the graph is plotted in Fig.~\ref{fig:pq}, 
at zero temperature and for $t=0.23$. \\
The transition from the uniform phase to the valence bond glass is continuous: 
the free energy of the two phases coincide within our numerical accuracy 
on the line $t_s(T)$ 
 where the liquid phase becomes unstable.  Close to
the transition point, the distribution of the $\chi_{ij}$ is peaked
around the value $\chi$ which characterizes the uniform phase, and it gets
broader and broader as the temperature and/or the hopping amplitude
are decreased. 
%The transition from the uniform phase to the VBG 
This transition
shares many common features with the transition from the paramagnetic phase
to the spin glass phase
observed in mean field (classical) spin glasses such as, for 
instance, the Sherrington-Kirkpatrick model~\cite{SG}: in both cases, one finds a continuous transition
with a continuous distribution of the overlaps. 
As a consequence it is natural to investigate whether the VBG phase is 
marginally stable as the spin glass phase \cite{SG}. This means that 
the VBG 
phase is critical not only at the transition but in the whole region of the
phase diagram where it exists.  
In order to do that we study whether the spatial correlations
among valence bonds on different links of the lattice 
$\langle \chi_{ij}(\omega_n) \, \chi_{kl}(\omega_n) \rangle_c^2$
are long-ranged 
(as previously we focus on $\omega_n=0$ which is expected to give the main
contribution). 
The inverse of the free energy Hessian matrix gives directly 
the dimer-dimer correlations. Instead of inverting this matrix, we follow   
a less computational demanding route using
a kind of fluctuation-dissipation relation. The idea is 
to measure the response of the system, more precisely of the value of 
$\chi_{ij}$,
to an external perturbation and relate it to the VBG susceptibility. 
The relevant perturbation for the present
case is a local increase of the hopping amplitude on a given link of the
graph, $t \to t + \delta t_{kl}$. 
Simple integrations by parts in the functional integral defining the partition function, Eq.~(\ref{eq:z}), allow one to establish the following identity:
\begin{equation} \label{chivbg}
\chi_{\textrm{VBG}} = \frac{1}{z {\cal N}} \! \sum_{(ij)\ne (kl)} \!\! 
\left \langle
\chi_{ij}^0 \, \chi_{kl}^0 \right \rangle_c^2 = \frac{1}{z {\cal N}} \!
\sum_{(ij)\ne (kl)} \! 
\left( \frac{ \textrm{d} \langle \chi_{ij}^0 \rangle}{\textrm{d} 
t_{kl}} \right)^2.
\end{equation}
where $\chi_{ij}^0$ is a short-hand notation for $\chi_{ij}(\omega_n=0)$ and 
the subscript $c$ denotes the connected correlation function.
We measured the response functions in the RHS of eq. (\ref{chivbg}).
We found, as shown in Fig.~\ref{fig:susc}, 
that the valence bond glass non-linear 
susceptibility, $\chi_{\textrm{VBG}}$, 
diverges as a power law both at fixed $t$ as 
the temperature is decreased ($\chi_2 \sim (T - T_s)^{-\gamma}$), 
and at fixed $T$ (included $T=0$) as the hopping is decreased ($\chi_2 \sim (t - t_s)^{-\gamma'}$).  
The exponents have the mean field value $\gamma \simeq \gamma' \simeq 1$.
Furthermore we find that $\chi_{\textrm{VBG}}$ is infinite (meaning of 
the order of, and scaling as, $\cal N$) in all the VBG phase, hence, confirming 
the marginality of the VBG phase.\\
Differently from the transition from the liquid phase to the VBG, the
transition from the dimer phase to the glassy one is discontinuous. It
takes place at $t_c(T)$, where the free energies of the two phases
coincide (at $T=0$ we have that $t_c \simeq 0.175$). The
dimer phase becomes unstable only for larger values of $t$. Furthermore 
the non-linear susceptibility, $\chi_{\textrm{VBG}}$, stays finite approaching
VBG from the dimer phase as it is expected for a first order 
transition.\\
\begin{figure}
\begin{center}
\includegraphics[scale=0.29,angle=270]{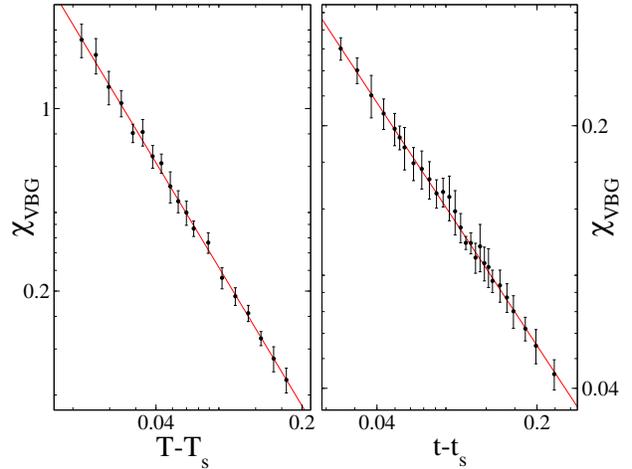}
\end{center} 
\vspace{-0.35cm}
\caption{Valence bond non-linear susceptibility, $\chi_{\textrm{VBG}}$, 
as a function
of $T-T_s$ at fixed $t = 0.1$ (left panel) and as a function of $t - t_s$
at zero temperature (right panel). 
$\chi_{\textrm{VBG}}$ 
diverges as a power law as the transition to the VBG is approached. 
In both cases the exponent is compatible with $\gamma \sim
1$. The data are averaged over $8$ different realization of graphs
with ${\cal N} = 512$ sites.} 
\vspace{-0.35cm}
\label{fig:susc}
\end{figure}
In summary the Valence Bond Glass phase 
is characterized by an amorphous arrangement of dimers 
and absence of magnetic ordering. 
It has long-range critical dimer-dimer correlations in the whole VBG phase 
(not only at the transition). 
It has no gap in the electronic and spin density of states, although 
we observe a pseudo-gap. As a consequence it is related to, 
but quite different from, valence bond crystal and spin glass phases.   
We expect the VBG phase to be 
generically one of the possible low temperature phases arising 
from the interplay of strong quantum fluctuations and frustration. 
In the future it would 
be important to go beyond the simplifying framework we focused on.
The role of $1/N$ corrections should be elucidated. Furthermore, 
it would be interesting to study different models, different 
(and more realistic) 
lattices and add some kind of local quenched disorder. 
The large $N$ approximation and the type of lattice we chose
favor the glassy phase. In reality we expect that VBG will emerge
as a true thermodynamic phase only in presence of some kind of quenched disorder
(not much if there is already geometrical frustration). 
In this case the VBG phase will be in competition with the spin glass phase
which in our treatment is excluded from the beginning because of the type of 
large $N$ limit we used. Another interesting route to follow is to study the 
effect of doping and the resulting properties of the VBG phase. This could
be relevant for underdoped high $T_c$ superconducting materials. Although 
the VBG phase may not be a true thermodynamic stable phase it could 
nevertheless capture some kind of metastable slow and glassy dynamics 
which seems indeed to be present~\cite{kohsaka}.
From a more fundamental and technical point of view 
obtaining a complete solution of our model 
(analytically or by numerical simulations) 
would be important 
to determine whether, as our results suggest, 
the VBG phase is completely analogous to the mean-field spin glass 
phase~\cite{SG}. Finally, it is worth studying the effect of 
magneto-elastic couplings. Because of the marginal stability of the VBG
phase they could play a very important role.         
We expect as experimental signature of the valence bond glass phase
spatially heterogeneous NMR signals.   
Furthermore, approaching the (continuous) transition toward the VBG phase,
the VBG susceptibility diverges and this could lead to 
anomalous (even divergent) non-linear pressure responses.     
Finally, we point out that preliminary results on modified random lattices 
(e.g., random regular graphs where each site is replaced by square plaquettes) 
show that also glassy flux phases \cite{affleck} might appear. These are 
characterized by amorphous circulating micro-currents. 
\acknowledgments 
We thank J.-P. Bouchaud, C. Chamon, A. Lef{\`e}vre, 
M. M{\'e}zard, G. Misguich and E. Vincent 
for many useful and helpful discussions. 

\section{Appendix} 
Here we describe in detail the 
Monte Carlo annealing simulations we used. 
We pick up a link $(ij)$ at random out of the $z {\cal N} /2$ total links 
and try to change either the real or the imaginary part of $\chi_{ij}$ by 
a random amount $\delta \in (-\delta_{max},\delta_{max})$ 
with probability $1/2$ respectively\footnote{Equivalently, at each step one can also attempt to 
change either the norm of the valence bond, $|\chi_{ij}|^2$, by an 
amount $\delta$, or its angle in the complex plane 
$\theta = \tan^{-1} [ \Im (\chi_{ij}) / \Re (\chi_{ij})]$, by randomly choosing 
a new angle $\theta$ in the interval $(0,2 \pi)$.}.
%We have verified that these two procedures 
%leads to the same results.}. 
Then we compute the new free energy, according to Eq.~(\ref{eq:free}).  
Since $F(\chi)$ contains a non-local term, 
at each step we have to diagonalize the 
matrix $t \mathbb{C}+\hat{\chi}$ and compute all its 
eigenvalues, which takes a computational time 
proportional to ${\cal N}^2$. 
The move is accepted with 
probability $ p = \textrm{min} \left \{ 1, \exp [ -\Delta F / 
T_{aux}] \right \}$. The value of $\delta_{max}$ is self-adapted during the 
simulation in such a way that the average acceptance rate of the moves 
is $0.3$. We have checked that several 
different values of the chosen acceptance rate lead to the same results.
The auxiliary temperature is decreased 
at constant rate down to very low temperature, starting from $T_{aux} = 0.5$. 
Most of the results presented 
here have been obtained with a rate $\Gamma = 
\dot{T}_{aux}/T_{aux} = 1.3 \cdot 10^{-3}$ 
(where each MC step consists of $z {\cal N}$ total attempts). 
We have verified that slower cooling rates 
down to $\Gamma \sim 5 \cdot 10^{-5}$ do not change the results. 
Some MC steps are finally performed at $T_{aux} = 0$.
%, where a move is accepted 
%only if it decreases the free energy. 
%In order to test our numerical method, we have performed some simulations on a 
%$10 \times 10$ square lattice, where we have been able to recover all the 
%results reported in~\cite{affleck}.

\end{document}